\documentclass[final,sort&compress,merge,3p]{elsarticle}
\usepackage{amsfonts}
\usepackage{amsthm}
\usepackage{amsmath}

\newcommand{\secref}[1]{Section~\ref{#1}}
\newcommand{\Secref}[1]{Section~\ref{#1}}
\newcommand{\subsecref}[1]{Subsection~\ref{#1}}

\newcommand{\braket}[1]{\langle{#1}\rangle}

\newcommand{\lshad}{[\![}
\newcommand{\rshad}{]\!]}
\newcommand{\sdot}{\,\cdot\,}
\newcommand{\dd}{\mathrm{d}}

\newcommand{\ud}{\,\dd}
\newcommand{\nbr}[1]{$#1$\nobreakdash-\hspace{0pt}}
\providecommand{\abs}[1]{\lvert#1\rvert}

\DeclareMathOperator{\id}{id}

\journal{arXiv}

\begin{document}

\begin{frontmatter}

\title{Quantum trajectories}
\author{Maciej B{\l}aszak}
\ead{blaszakm@amu.edu.pl}
\author{Ziemowit Doma{\'n}ski}
\ead{ziemowit@amu.edu.pl}
\address{Faculty of Physics, Adam Mickiewicz University, Pozna{\'n}, Poland}

\begin{abstract}
This paper presents a new approach to phase space trajectories in
quantum mechanics. A Moyal description of quantum theory is used,
where observables and states are treated as classical functions on
a classical phase space. A quantum trajectory being an appropriate
solution to quantum Hamiltonian equations is also a function
defined on a classical phase space. It results in a deformation of a classical action
of a flow on observables and states to an appropriate quantum
action. It also leads to a new multiplication rule
for any quantum trajectory treated as a one-parameter group of
diffeomorphisms. Moreover, several examples are given, presenting
the developed formalism for particular quantum systems.
\end{abstract}


\end{frontmatter}

\section{Introduction}
\label{sec:1}
The time evolution of a classical Hamiltonian system is fully given by
trajectories (a flow) in a phase space. Having calculated a classical flow
$\Phi_t$ for the given system a time evolution of states and observables can be
received by simply composing them with $\Phi_t$. A classical flow is defined as
a map $\Phi_t \colon M \to M$ on the phase space $M$, which at every point
$\xi_0 \in M$ gives a trajectory (curve) $\gamma(t) = \Phi_t(\xi_0)$ on $M$
passing through the point $\xi_0$ and being a solution of the Hamilton's
equations. Geometrically trajectories constitute a flow of a Hamiltonian vector
field. Furthermore, any trajectory $\Phi_t(\xi_0)$ has the property of being a
canonical transformation for every $t$, and the set
$\{\Phi_t\}_{t \in \mathbb{R}}$ have a structure of a group with multiplication
being a composition of maps.

From the very beginning of quantum physics, efforts have been taken to formulate
some kind of an analogue of phase space trajectories in quantum mechanics
\cite{Dirac:1958}. The most common approaches to quantum dynamics are the
de~Broglie-Bohm approach \cite{Broglie:1926a,*Broglie:1927b,Bohm:1952a,%
*Bohm:1952b,Holland:1993}, the average value approach
\cite{Wyatt:2005,Littlejohn:1986}, and the Moyal trajectories approach (see
\cite{Dias:2007,Krivoruchenko:2007} and references therein). Worth noting is
also the paper \cite{Rieffel:1997} written by Rieffel where he considers a
classical limit of a quantum time evolution in the framework of a strict
deformation quantization. In the following paper we develop the Moyal approach
to time evolution. The usual formulation of the theory of Moyal
trajectories is based on the phase space description of quantum mechanics, where
one considers the Heisenberg evolution of fundamental observables of position
and momentum, being \nbr{\hbar}deformation of the classical Hamiltonian
evolution. Moreover, the deformation of arbitrary order can be calculated by an
\nbr{\hbar}hierarchy of recursive first order linear partial differential
equations \cite{Osborn:1995,Dias:2007,Krivoruchenko:2007}. The time evolution
of observables cannot be given as a simple composition of observables with a
quantum flow. For this reason \citet{Dias:2007}, and \citet{Krivoruchenko:2007}
considered observables as \nbr{\star}functions and a quantum phase space as a
plane of noncommuting variables. Then the action of a flow on observables was
given as a \nbr{\star}composition.

In our approach to quantum trajectories we treat observables as ordinary
functions on a classical phase space. We present in explicit form a quantum action of a flow on
observables, which is a deformation of the respective classical action.
The resulting time dependence of observables gives an appropriate solution
of a quantum time evolution equation for observables \eqref{eq:2.2} (Heisenberg's representation on a phase space).
Then, we show that a set of quantum symplectomorphisms (quantum flow) has a structure of a group with multiplication
(quantum composition) being a deformation of the ordinary composition considered as a multiplication in a group of classical symplectomorphisms (classical flow). The explicit form of the quantum composition law is presented. Such
approach to quantum trajectories have a benefit in that it is not needed to
calculate the form of observables as \nbr{\star}functions, but only a
quantum action of a given trajectory needs to be found. Also we expect
that our approach to quantum flows will allow a development of phase space
quantum mechanics in a geometrical setting similar to that of classical
Hamiltonian mechanics.

The paper is organized as follows. In \secref{sec:2} we review a basics of
a quantum mechanics on a phase space. \Secref{sec:3} contains the main
results of a theory of quantum trajectories. Finally, in \secref{sec:4}, examples of
particular quantum systems are presented.

\section{Phase space quantum mechanics}
\label{sec:2}

\subsection{Preliminaries}
\label{subsec:2.1}
The most natural approach to quantum theory, when dealing with quantum
trajectories, is a phase space quantum mechanics (see \cite{Bayen:1978a,%
*Bayen:1978b}, and \cite{Dito.Sternheimer:2002,Gutt:2000,Weinstein:1994b} for
recent reviews). The following review of phase space quantum mechanics
comes from \cite{Blaszak:2012}. The phase space approach to
quantum theory is based on an appropriate deformation of a
classical Hamiltonian mechanics, with respect to some parameter
which we take to be the Planck's constant $\hbar$. The deformation
of a classical Hamiltonian system can be fully given by deforming
an algebraic structure of a classical Poisson algebra. This will
then yield a deformation of a phase space (a Poisson manifold) to
a noncommutative phase space (a noncommutative Poisson manifold),
a deformation of classical states to quantum states and a
deformation of classical observables to quantum observables.

First, let us deal with a deformation of a phase space. A Poisson
manifold $(M,\mathcal{P})$ ($\mathcal{P}$ being a Poisson tensor)
is fully described by a Poisson algebra $\mathcal{A}_C =
(C^\infty(M),\cdot,\{\sdot,\sdot\})$ of smooth complex-valued
functions on the phase space $M$, where $\cdot$ is a point-wise
product of functions and $\{\sdot,\sdot\}$ is a Poisson bracket
induced by a Poisson tensor $\mathcal{P}$. Hence by deforming
$\mathcal{A}_C$ to some noncommutative algebra $\mathcal{A}_Q =
(C^\infty(M),\star,\lshad\sdot,\sdot\rshad)$, where $\star$ is
some noncommutative associative product of functions being a
deformation of a point-wise product, and $\lshad\sdot,\sdot\rshad$
is a Lie bracket satisfying the Leibniz's rule and being a
deformation of the Poisson bracket $\{\sdot,\sdot\}$, we can think
of a quantum Poisson algebra $\mathcal{A}_Q$ as describing a
noncommutative Poisson manifold.

The algebra $\mathcal{A}_C$ contains in particular a subset of
classical observables, whereas $\mathcal{A}_Q$ contains a subset
of quantum observables. Note that quantum observables are
functions on the phase space $M$ similarly as in classical
mechanics. Furthermore, classical observables are real-valued
functions from $\mathcal{A}_C$, i.e., self-adjoint functions with
respect to the complex-conjugation --- an involution in the
algebra $\mathcal{A}_C$. Quantum observables should also be
self-adjoint functions with respect to an involution in the
algebra $\mathcal{A}_Q$. However, in general the
complex-conjugation do not need to be an involution in
$\mathcal{A}_Q$. Thus in $\mathcal{A}_Q$ we have to introduce some
involution which would be a deformation of the complex-conjugation
\cite{Blaszak:2012}. As a consequence, quantum observables
(self-adjoint functions with respect to the quantum involution)
might be complex and $\hbar$-dependent.

There is a vast number of equivalent quantization schemes (see
\cite{Blaszak:2012} for review of the subject) which yield a
quantization equivalent to a standard approach to quantum
mechanics but giving different orderings of position and momentum
operators. From this diversity of quantization schemes the
simplest one is a Moyal quantization. It follows from the fact
that for the Moyal quantum algebra the involution is the
complex-conjugation as in the classical case. Thus in this case
quantum observables, exactly like classical observables, can be
chosen as real-valued functions. Further on we will deal only with
that distinguished quantization. Such a choice is not a
restriction as other quantization schemes known in the literature
are gauge equivalent to the Moyal one (see \cite{Blaszak:2012} and
\subsecref{subsec:2.4}).

The Moyal quantization scheme is as follows. First, let us assume that
$M = \mathbb{R}^{2N}$ and $\mathcal{P} = \partial_{x^i} \wedge \partial_{p_i}$.
Define a \nbr{\star}product by a formula
\begin{equation*}
f \star g = f \exp \left( \frac{1}{2} i\hbar \overleftarrow{\partial}_{x^i}
    \overrightarrow{\partial}_{p_i} - \frac{1}{2} i\hbar
    \overleftarrow{\partial}_{p_i} \overrightarrow{\partial}_{x^i} \right) g.
\end{equation*}
This \nbr{\star}product is called the Moyal product.
For a two-dimensional case ($N = 1$) the Moyal product reads
\begin{equation*}
f \star g = \sum_{k=0}^\infty \frac{1}{k!}
\left(\frac{i\hbar}{2}\right)^{k}
    \sum_{m=0}^k \binom{k}{m} (-1)^m
    (\partial_x^{k-m}\partial_p^m f)(\partial_x^m\partial_p^{k-m} g).
\end{equation*}
The deformed Poisson bracket $\lshad\sdot,\sdot\rshad$ associated with the
\nbr{\star}product will be given in terms of a \nbr{\star}commutator
$[\sdot,\sdot]$ as follows
\begin{equation*}
\lshad f,g \rshad = \frac{1}{i\hbar} [f,g]
= \frac{1}{i\hbar}(f \star g - g \star f), \quad
f,g \in \mathcal{A}_Q.
\end{equation*}

To avoid problems with convergence of the series in the above definition of the
\nbr{\star}product the common practice is to extend the space $C^\infty(M)$ to
a space $C^\infty(M)\lshad\hbar\rshad$ of formal power series in $\hbar$ with
coefficients from $C^\infty(M)$. The \nbr{\star}product is then properly defined
on such space.

Observe that every function $f \in \mathcal{A}_Q$ can be expanded into a
\nbr{\star}power series
\begin{equation*}
f = \sum_{n,m = 0}^\infty a_{nm}
    \underbrace{x \star \dotsb \star x}_n \star
    \underbrace{p \star \dotsb \star p}_m,
\end{equation*}
where $a_{nm} \in \mathbb{C}$. Indeed, the result follows from the fact that
every monomial $x^n p^m$ can be written as a \nbr{\star}polynomial, which on the
other hand can be seen from the definition of the \nbr{\star}product.

\subsection{Space of states, expectation values of observables and time
evolution equation}
\label{subsec:2.2}
In general a space of states is fully characterized by the algebraic structure
of the quantum Poisson algebra $\mathcal{A}_Q$ \cite{Bordemann:1998,%
Waldmann:2005}. It can be shown that for the Moyal quantization states can
be represented as quantum distribution functions, i.e., square integrable
functions $\rho$ defined on the phase space satisfying certain conditions
\cite{Dias:2004b,Blaszak:2012}. For this reason the Hilbert space $\mathcal{H} =
L^2(\mathbb{R}^{2N})$ of square integrable functions on the phase space will be
called a space of states. Observe, that the Moyal product can be extended to a
product between smooth functions from $C^\infty(\mathbb{R}^{2N})$ and square
integrable functions from $L^2(\mathbb{R}^{2N})$.

Formulas for the expectation values of observables and the time evolution of
states are similar as in classical mechanic, except that the point-wise product
$\cdot$ of functions and the Poisson bracket $\{\sdot,\sdot\}$ have to be
replaced with the \nbr{\star}product and the deformed Poisson bracket
$\lshad\sdot,\sdot\rshad$. Thus the expectation value of an observable
$A \in \mathcal{A}_Q$ in a state $\rho \in L^2(\mathbb{R}^{2N})$ is given by
the formula
\begin{align*}
\braket{A}_{\rho} & = \iint (A(0) \star \rho(t))(x,p) \ud{x} \ud{p}
= \braket{A(0)}_{\rho(t)} \\
& = \iint (A(t) \star \rho(0))(x,p) \ud{x} \ud{p}
= \braket{A(t)}_{\rho(0)}.
\end{align*}
The time evolution equation of quantum distribution functions $\rho(t)$
(Schr{\"o}dinger picture) is the counterpart of the Liouville's equation
describing the time evolution of classical distribution functions, and is given
by the formula
\begin{equation*}
\frac{\dd \rho}{\dd t} - \lshad H,\rho(t) \rshad = 0
\quad \Leftrightarrow \quad
i\hbar \frac{\dd \rho}{\dd t} - [H,\rho(t)] = 0,
\end{equation*}
where $H$ is a Hamiltonian (distinguished observable from
$\mathcal{A}_Q$). The time evolution of quantum observable $A(t)$
(Heisenberg picture) is given by
\begin{equation}
\frac{\dd A}{\dd t} - \lshad A(t),H \rshad = 0
\quad \Leftrightarrow \quad
i\hbar \frac{\dd A}{\dd t} - [A(t),H] = 0.
\label{eq:2.2}
\end{equation}

\subsection{Equivalence of quantizations}
\label{subsec:2.4}
Two star-products $\star$ and $\star'$ are said to be gauge equivalent if there
exists a vector space automorphism $S \colon C^\infty(\mathbb{R}^{2N}) \to
C^\infty(\mathbb{R}^{2N})$ of the form
\begin{equation}
S = \sum_{k=0}^\infty \hbar^k S_k, \quad S_0 = 1,
\label{eq:2.3}
\end{equation}
where $S_k$ are linear operators, which satisfies the formula
\begin{equation*}
S(f \star g) = Sf \star' Sg, \quad f,g \in C^\infty(\mathbb{R}^{2N}).
\end{equation*}
If, moreover, the automorphism $S$ preserves the deformed Poisson brackets and
involutions $*$ and $*'$ from the algebras $\mathcal{A}_Q =
(C^\infty(\mathbb{R}^{2N}), \star, \lshad\sdot,\sdot\rshad, *)$ and
$\mathcal{A}'_Q = (C^\infty(\mathbb{R}^{2N}), \star', \lshad\sdot,\sdot\rshad',
*')$, i.e.,
\begin{equation*}
S(\lshad f,g \rshad) = \lshad Sf,Sg \rshad', \quad  S(f^*) = (Sf)^{*'},
\end{equation*}
then $S$ is an isomorphism of the algebra $\mathcal{A}_Q$ onto the algebra
$\mathcal{A}'_Q$.

Two quantizations of a classical Hamiltonian system are equivalent
if there exists an isomorphism $S$ of their quantum Poisson
algebras. This equivalence is mathematical as well as physical. It
has been stressed out in \subsecref{subsec:2.1}, that to the same
measurable quantity correspond different functions from respective
quantum Poisson algebras. This observation seems to be missing in
considerations of different quantizations present in the
literature. In fact, to every observable $A \in \mathcal{A}_Q$
from one quantization scheme corresponds an observable $A' = SA
\in \mathcal{A}'_Q$ from the other quantization scheme. Both
observables $A$ and $A'$ describe the same measurable quantity and
in the limit $\hbar \to 0$ reduce to the same classical
observable. Such approach to equivalence of quantum systems
introduces, indeed, physically equivalent quantizations as the
functions $A$, $A'$ from different quantization schemes have the
same spectra, expectation values, etc., and when they are
Hamiltonians they describe the same time evolution.

It is possible to define a morphism of spaces of states of different
quantization schemes, in terms of $S$. This morphism we will also denote by $S$.
In case when the initial quantization is the Moyal quantization $S$ will be a
Hilbert space isomorphism. In what follows we will restrict to the case when the
\nbr{S}image of the space of states $L^2(\mathbb{R}^{2N})$ is also a Hilbert
space $L^2(\mathbb{R}^{2N},\mu)$ of square integrable functions possibly with
respect to a different measure $\mu$.

\section{Quantum trajectories in phase space}
\label{sec:3}
As before we will consider the Moyal quantization of a classical Hamiltonian
system $(M,\mathcal{P},H)$, where $M = \mathbb{R}^{2N}$, $\mathcal{P} =
\partial_{x^i} \wedge \partial_{p_i}$, and $H \in C^\infty(M)$ is an arbitrary
real function.

The solution of quantum Hamiltonian equations
\begin{equation}
\dot{Q}^i(t) = \lshad Q^i(t),H \rshad, \quad
\dot{P}_j(t) = \lshad P_j(t),H \rshad,
\label{eq:3.4}
\end{equation}
where $Q^i(x,p,0) = x^i$ and $P_j(x,p,0) = p_j$, i.e., the Heisenberg
representation \eqref{eq:2.2} for observables of position and momentum,
generates a quantum flow $\Phi_t$ in phase space according to an equation
\begin{equation}
\Phi_t(x,p;\hbar) = (Q(x,p,t;\hbar),P(x,p,t;\hbar)).
\label{eq:3.5}
\end{equation}
For every instance of time $t$ the map $\Phi_t$ is a quantum canonical
transformation (quantum symplectomorphism) from coordinates $x,p$ to new
coordinates $x' = Q(x,p,t;\hbar), p' = P(x,p,t;\hbar)$. In other words $\Phi_t$
preserves the quantum Poisson bracket: $\lshad Q^i(t),P_j(t) \rshad =
\delta^i_j$ (this can be easily seen from \eqref{eq:3.6} and the fact that
$\lshad Q^i(0),P_j(0) \rshad = \lshad x^i,p_j \rshad = \delta^i_j$).

The flow $\Phi_t$, as every other quantum canonical
transformation, can act on observables and states as simple
composition of maps. Such classical action can also be used to
transform the algebraic structure of the quantum Poisson algebra
so that the action will be an isomorphism of the initial algebra
and its transformation. A star-product $\star_t$ being the Moyal
product transformed by $\Phi_t$ is defined by the formula
\begin{equation}
(f \star g) \circ \Phi_t^{-1} = (f \circ \Phi_t^{-1}) \star_t
    (g \circ \Phi_t^{-1}), \quad f,g \in C^\infty(\mathbb{R}^{2N}).
\label{eq:3.1}
\end{equation}
The \nbr{\star_t}product takes the form of the Moyal product but with
derivatives $\partial_{x^i}$, $\partial_{p_i}$ replaced by some other
derivations $D_{x^i}$, $D_{p_i}$ of the algebra $C^\infty(\mathbb{R}^{2N})$:
\begin{equation*}
f \star_t g = f \exp \left( \frac{1}{2} i\hbar \overleftarrow{D_{x^i}}
    \overrightarrow{D_{p_i}} - \frac{1}{2} i\hbar \overleftarrow{D_{p_i}}
    \overrightarrow{D_{x^i}} \right) g,
\end{equation*}
where derivations $D_{x^i}$, $D_{p_i}$ are transformations of the derivatives
$\partial_{x^i}$, $\partial_{p_i}$:
\begin{equation*}
(\partial_{x^i} f) \circ \Phi_t^{-1} = D_{x^i}(f \circ \Phi_t^{-1}), \quad
(\partial_{p_i} f) \circ \Phi_t^{-1} = D_{p_i}(f \circ \Phi_t^{-1}).
\end{equation*}
The \nbr{\star_t}product can be also written in a different form, a so called
covariant form. For more details see e.g. \cite{Bayen:1978a,Dias:2001,%
Dias:2004}.
The crucial point of our construction is the observation that for a wide class
of quantum flows the \nbr{\star_t}product is gauge equivalent to the Moyal
product. Strictly speaking, to a quantum flow $\Phi_t$ there corresponds a
unique isomorphism $S_t$ of the form \eqref{eq:2.3} satisfying
\begin{subequations}
\label{eq:3.2}
\begin{gather}
S_t(f \star g) = S_t f \star_t S_t g, \label{eq:3.2a} \\
S_t x^i = x^i, \quad S_t p_j = p_j, \label{eq:3.2b} \\
S_t(f^*) = (S_t f)^*. \label{eq:3.2c}
\end{gather}
\end{subequations}
We will consider only such flows to which an isomorphism $S_t$ can be
associated, however, we believe that this holds for every quantum flow. Note,
that for the \nbr{\star_t}algebra the involution is also the
complex-conjugation.

A formal solution of the time evolution equation \eqref{eq:2.2} for an
observable $A \in \mathcal{A}_Q$ can be expressed by the formula
\begin{equation*}
A(t) = e^{-t\lshad H,\sdot \rshad} A(0)
= e_\star^{\frac{i}{\hbar}tH} \star A(0) \star e_\star^{-\frac{i}{\hbar}tH},
\end{equation*}
where
\begin{equation*}
e^{-t\lshad H,\sdot \rshad} := \sum_{k=0}^{\infty} \frac{1}{k!} (-t)^k
\underbrace{\lshad H,\lshad H,\dotsc \lshad H,\sdot \rshad \dotsc \rshad
\rshad}_k
\end{equation*}
and
\begin{equation*}
e^{\frac{i}{\hbar}tH}_\star := \sum_{k=0}^{\infty} \frac{1}{k!}
\left(\frac{i}{\hbar}t\right)^k \underbrace{H \star \dotsb \star H}_k.
\end{equation*}
In particular, the solution of \eqref{eq:3.4} takes the form
\begin{subequations}
\label{eq:3.6}
\begin{align}
Q^i(t) & = e^{-t\lshad H,\sdot \rshad} Q^i(0)
= e_\star^{\frac{i}{\hbar}tH} \star Q^i(0) \star e_\star^{-\frac{i}{\hbar}tH},
\label{eq:3.6a} \\
P_j(t) & = e^{-t\lshad H,\sdot \rshad} P_j(0)
= e_\star^{\frac{i}{\hbar}tH} \star P_j(0) \star e_\star^{-\frac{i}{\hbar}tH},
\label{eq:3.6b}
\end{align}
\end{subequations}
which for fixed initial condition $Q^i(x,p,0) = x^i$ and $P_j(x,p,0) = p_j$
represents a particular quantum trajectory.

A time evolution of an observable $A \in \mathcal{A}_Q$ should be alternatively
expressed by action of the quantum flow $\Phi_t$ on $A$. The composition of
$\Phi_t$ with observables (the classical action of $\Phi_t$ on observables) does
not result in a proper time evolution of observables. Thus it is necessary to
deform this classical action. We will prove that a proper action of the quantum
flow $\Phi_t$ on functions from $\mathcal{A}_Q$ (a pull-back of $\Phi_t$) is
given by the formula
\begin{equation}
\Phi_t^* A = (S_t A) \circ \Phi_t,
\label{eq:3.7}
\end{equation}
where $S_t$ is an isomorphism associated to the quantum canonical transformation
$\Phi_t^{-1}$. Indeed, \eqref{eq:3.7} can be proved first by noting that
\begin{equation*}
\Phi_t^* Q^i(0) = (S_t Q^i(0)) \circ \Phi_t = Q^i(0) \circ \Phi_t = Q^i(t)
= e^{-t\lshad H,\sdot \rshad} Q^i(0)
\end{equation*}
and similarly
\begin{equation*}
\Phi_t^* P_j(0) = e^{-t\lshad H,\sdot \rshad} P_j(0),
\end{equation*}
where the fact that $S_t x^i = x^i$ and $S_t p_j = p_j$ was used,
which on the other hand was a consequence of the quantum
canonicity of $\Phi_t$. Secondly, $\Phi_t^*$ given by
\eqref{eq:3.7} is an automorphism of $\mathcal{A}_Q$ as
\begin{equation*}
\Phi_t^*(A \star B) = (S_t(A \star B)) \circ \Phi_t
= (S_t A \star_t S_t B) \circ \Phi_t
= \left((S_t A) \circ \Phi_t\right) \star \left((S_t B) \circ \Phi_t\right)
= \Phi_t^*A \star \Phi_t^*B,
\end{equation*}
where $\star_t$ denotes a star-product transformed by $\Phi_t^{-1}$. Thus
\begin{equation}
\Phi_t^* = e^{-t\lshad H,\sdot \rshad}
\label{eq:3.8}
\end{equation}
holds true since every function from $\mathcal{A}_Q$ can be expressed as a
\nbr{\star}power series (see \subsecref{subsec:2.1}).

In a complete analogy with classical theory one can define a quantum Hamiltonian
vector field by $\zeta_H = \lshad \sdot,H \rshad$. Then \eqref{eq:3.8} states
that $\Phi_t$ is a flow of the quantum Hamiltonian vector field $\zeta_H$. Also
in an analogy with classical mechanics $\{\Phi_t\}$ is a one-parameter group of
quantum canonical transformations with respect to a multiplication defined by
\begin{equation}
\Phi_{t_1} \Phi_{t_2} = (S_{t_2} \Phi_{t_1}) \circ \Phi_{t_2},
\label{eq:3.15}
\end{equation}
where $S_{t_2} \Phi_{t_1}$ denotes a map $\mathbb{R}^{2N} \to \mathbb{R}^{2N}$
given by the formula
\begin{equation*}
S_{t_2} \Phi_{t_1} = (S_{t_2}Q^1(t_1),\dotsc,S_{t_2}P_N(t_1)),
\end{equation*}
where $\Phi_{t_1} = (Q^1(t_1),\dotsc,Q^N(t_1),P_1(t_1),\dotsc,P_N(t_1))$.
Multiplication defined in such a way satisfies properties similar to their
classical counterparts:
\begin{equation*}
\Phi_0 = \id, \quad \Phi_{t_1} \Phi_{t_2} = \Phi_{t_1 + t_2},
\end{equation*}
proving that $\{\Phi_t\}$ is a group. Further on we will call it a quantum composition. The quantum composition rule given by
\eqref{eq:3.15} is properly defined since it respects the quantum pull-back of
flows:
\begin{equation}
(\Phi_{t_1} \Phi_{t_2})^* = \Phi_{t_2}^* \circ \Phi_{t_1}^*.
\label{eq:3.14}
\end{equation}
Indeed, it is enough to show \eqref{eq:3.14} for an arbitrary
\nbr{\star}monomial. For simplicity we will present the proof for a
two-dimensional case and for a \nbr{\star}monomial $x \star p$. Using the fact
that $S_t x = x$ and $S_t p = p$ for every $t$, following from quantum
canonicity of the flow $\Phi_t$, one calculates that
\begin{align*}
(\Phi_{t_2}^* \circ \Phi_{t_1}^*)(x \star p) & =
    \Phi_{t_2}^*\bigl( (S_{t_1}(x \star p)) \circ \Phi_{t_1} \bigr)
= \Phi_{t_2}^* \bigl( (x \star_{t_1} p) \circ \Phi_{t_1} \bigr)
= \Phi_{t_2}^* \bigl( Q(t_1) \star P(t_1) \bigr) \\
& = \bigl( S_{t_2}(Q(t_1) \star P(t_1)) \bigr) \circ \Phi_{t_2}
= \bigl( S_{t_2} Q(t_1) \star_{t_2} S_{t_2} P(t_1) \bigr) \circ \Phi_{t_2} \\
& = (x \star_{t_2,t_1} p) \circ S_{t_2} \Phi_{t_1} \circ \Phi_{t_2},
\end{align*}
where $\star_{t_1}$, $\star_{t_2}$, denote Moyal products transformed,
respectively, by transformations $\Phi_{t_1}^{-1}$, $\Phi_{t_2}^{-1}$, and
$\star_{t_2,t_1}$ denotes the \nbr{\star_{t_2}}product transformed by
$(S_{t_2}\Phi_{t_1})^{-1}$. Now, from the relation $S_{T_1 \circ T_2} =
S_{T_1,T_2} S_{T_1}$ valid for any transformations $T_1$, $T_2$ defined on the
whole phase space ($S_{T_1 \circ T_2}$ is an isomorphism intertwining
star-products $\star$ and $\star_{T_1 \circ T_2}$, $S_{T_1,T_2}$ intertwines
$\star_{T_1}$ with $\star_{T_1 \circ T_2}$, and $S_{T_1}$ intertwines $\star$
with $\star_{T_1}$, where $\star_{T_1}$ and $\star_{T_1 \circ T_2}$ are Moyal
products transformed, respectively, by transformations $T_1$ and
$T_1 \circ T_2$), one receives that
\begin{equation*}
S_{(\Phi_{t_1} \Phi_{t_2})^{-1}}(x \star p) =
    S_{\Phi_{t_2}^{-1},(S_{t_2}\Phi_{t_1})^{-1}} S_{t_2}(x \star p)
= S_{\Phi_{t_2}^{-1},(S_{t_2}\Phi_{t_1})^{-1}}(x \star_{t_2} p)
= x \star_{t_2,t_1} p.
\end{equation*}
Hence
\begin{equation*}
(\Phi_{t_2}^* \circ \Phi_{t_1}^*)(x \star p) = S_{(\Phi_{t_1}\Phi_{t_2})^{-1}}
    (x \star p) \circ S_{t_2}\Phi_{t_1} \circ \Phi_{t_2}
= (\Phi_{t_1} \Phi_{t_2})^*(x \star p).
\end{equation*}

In the limit $\hbar \to 0$, \eqref{eq:3.6} reduces to classical phase space
trajectories
\begin{gather*}
Q^i(t) = e^{-t \{H,\sdot\}} Q^i(0), \quad
P_j(t) = e^{-t \{H,\sdot\}} P_j(0), \\
Q^i(x,p,0) = x^i, \quad P_j(x,p,0) = p_j,
\end{gather*}
which are formal solutions of classical Hamiltonian equations
\begin{equation}
\dot{Q}^i(t) = \{ Q^i(t),H \}, \quad \dot{P}_j(t) = \{ P_j(t),H \}.
\label{eq:3.9}
\end{equation}
In more explicit form classical trajectories are represented by a flow
(diffeomorphism)
\begin{equation}
\Phi_t(x,p) = (Q(x,p,t),P(x,p,t)),
\label{eq:3.10}
\end{equation}
which is an $\hbar \to 0$ limit of the quantum flow \eqref{eq:3.5}.
Diffeomorphism \eqref{eq:3.10} is a classical symplectomorphism. An action of
the classical flow $\Phi_t$ on functions from $\mathcal{A}_C$ (a pull-back of
$\Phi_t$) is just a simple composition of functions with $\Phi_t$, being an
$\hbar \to 0$ limit of \eqref{eq:3.7}
\begin{equation}
\Phi_t^* A = A \circ \Phi_t.
\label{eq:3.11}
\end{equation}
$\{\Phi_t\}$ forms a one-parameter group
of canonical transformations, preserving a classical Poisson bracket:
$\{Q^i(t),P_j(t)\} = \delta^i_j$, with a multiplication being an ordinary
composition of maps
\begin{equation}
\Phi_{t_1} \Phi_{t_2} = \Phi_{t_1} \circ \Phi_{t_2},
\label{eq:3.12}
\end{equation}
which is the $\hbar \to 0$ limit of \eqref{eq:3.15}.

\section{Examples}
\label{sec:4}

\subsection{Example 1: Harmonic oscillator}
\label{subsubsec:4.1}
In this example we will consider quantum trajectories of the harmonic
oscillator. The Hamiltonian of the harmonic oscillator is given by the equation
\begin{equation*}
H(x,p) = \frac{1}{2} \left( p^2 + \omega^2 x^2 \right).
\end{equation*}
It happens that in such case the quantum trajectory coincides with
the classical one. Indeed, one can show that
\begin{align*}
Q(t) & = e^{-t\lshad H,\sdot \rshad} Q(0) = e^{-t\{H,\sdot\}} Q(0), \\
P(t) & = e^{-t\lshad H,\sdot \rshad} P(0) = e^{-t\{H,\sdot\}} P(0)
\end{align*}
and in explicit form classical/quantum trajectory $\Phi_t =
(Q(t),P(t))$ of a harmonic oscillator is
\begin{align*}
Q(x,p,t) & = x \cos \omega t + \omega^{-1} p \sin \omega t, \\
P(x,p,t) & = p \cos \omega t - \omega x \sin \omega t.
\end{align*}

Observe that the classical action (composition) of $\Phi_t$ on the algebra of
observables preserves the Moyal product, i.e.,
\begin{equation*}
(f \star g) \circ \Phi_t = (f \circ \Phi_t) \star (g \circ \Phi_t), \quad
f,g \in C^\infty(\mathbb{R}^{2N}).
\end{equation*}
Thus in accordance with \eqref{eq:3.2} the unique isomorphism $S_t$ associated
with $\Phi_t$ is equal $S_t = 1$. This means that the action of the flow
$\Phi_t$ on observables \eqref{eq:3.7} as well as the quantum composition rule
\eqref{eq:3.15} for the flow is equal to the classical composition rule of that
flow. In other words the time evolution of observables is the same as in
classical case. The difference between the classical and quantum system is in
the admissible states which evolve along the flow. In classical case states are
probabilistic distribution functions, whereas in quantum case states are
quasi-probabilistic distribution functions. In particular, classical pure states
are Dirac distribution functions; however, quantum pure states will no longer be
of such form due to the Heisenberg uncertainty principle.

\subsection{Example 2}
\label{subsec:4.2}
In this example let us consider a two particle system described by the
Hamiltonian
\begin{equation*}
H(x,p) = \frac{p_1^2}{2m_1} + \frac{p_2^2}{2m_2} + k x^1 p_2^2,
\end{equation*}
where $m_1$ and $m_2$ are masses of particles and $k$ is a coupling constant.
The solution of quantum Hamiltonian equations \eqref{eq:3.4} reads
\cite{Dias:2004}
\begin{align*}
Q^1(t) & = x^1 + \frac{1}{m_1} p_1 t - \frac{k}{2m_1} p_2^2 t^2,
\displaybreak[0] \\
P_1(t) & = p_1 - k p_2^2 t, \displaybreak[0] \\
Q^2(t) & = x^2 + \left( \frac{1}{m_2} p_2 + 2k x^1 p_2 \right) t
    + \frac{k}{m_1} p_1 p_2 t^2 - \frac{k^2}{3m_1} p_2^3 t^3,
\displaybreak[0] \\
P_2(t) & = p_2,
\end{align*}
which coincides again with a solution of classical Hamiltonian
equations. However, in accordance with \eqref{eq:3.1} the received quantum flow
$\Phi_t$ transforms the Moyal product to the following product
\begin{equation*}
f \star_t g = f \exp \left( \frac{1}{2} i\hbar \overleftarrow{D_{x^i}}
    \overrightarrow{D_{p_i}} - \frac{1}{2} i\hbar \overleftarrow{D_{p_i}}
    \overrightarrow{D_{x^i}} \right) g,
\end{equation*}
where
\begin{align*}
D_{x^1} & = \partial_{x^1} + 2kt p_2 \partial_{x^2}, \\
D_{p_1} & = \partial_{p_1} + \frac{1}{m_1}t \partial_{x^1}
    + \frac{k}{m_1}t^2 p_2 \partial_{x^2}, \\
D_{x^2} & = \partial_{x^2}, \\
D_{p_2} & = \partial_{p_2} - 2kt p_2 \partial_{p_1}
    - \frac{k}{m_1}t^2 p_2 \partial_{x^1}
    + \left( \frac{1}{m_2}t + 2kt x^1
    - \frac{k}{m_1}t^2 p_1 - \frac{k^2}{m_1}t^3 p_2^2 \right) \partial_{x^2}.
\end{align*}
Moreover, the isomorphism $S_t$ associated with $\Phi_t$ and intertwining the
Moyal product with the \nbr{\star_t}product takes the form
\begin{equation*}
S_t = \exp\left(
    \frac{1}{8} \hbar^2 \frac{k}{m_1} t^2 \partial_{x^1} \partial_{x^2}^2
    + \frac{1}{4} \hbar^2 kt \partial_{p_1} \partial_{x^2}^2
    + \frac{1}{12} \hbar^2 \frac{k^2}{m_1} t^3 p_2 \partial_{x^2}^3 \right).
\end{equation*}
Indeed, a direct calculations show that the relations \eqref{eq:3.2} are
satisfied. More details of the construction of $S_t$ the reader can find in
\cite{Blaszak:2012c}.

As in this case $S_{t_2} \Phi_{t_1}=\Phi_{t_1}$, the group multiplication for $\{\Phi_t\}$ is just a
composition of maps, as one could expect since $\Phi_t$ is simultaneously the
classical and quantum trajectory. However, the action of $\Phi_t$ on observables
and states does not reduce in general to a composition of maps \eqref{eq:3.11}. This shows that
the time evolution of quantum observables differs in general from the time
evolution of classical observables.

One can check by direct calculations that the action of the quantum flow
$\Phi_t$ on an observable $A$, given by \eqref{eq:3.7}, indeed describes the
quantum time evolution of $A$. As an example let us take $A(x,p) = x_1 x_2^2$.
Then
\begin{equation*}
(S_t A)(x,p) = x_1 x_2^2 + \frac{1}{4} \hbar^2 \frac{k}{m_1} t^2
\end{equation*}
and it can be checked that
\begin{equation*}
A(t) = (S_t A) \circ \Phi_t = Q^1(t)(Q^2(t))^2
    + \frac{1}{4} \hbar^2 \frac{k}{m_1} t^2
\end{equation*}
satisfies the time evolution equation \eqref{eq:2.2}.

\subsection{Example 3}
\label{subsec:4.3}
In this example we will consider a system described by a Hamiltonian
\begin{equation*}
H(x,p) = x^2 p^2.
\end{equation*}
The solution of quantum Hamiltonian equations \eqref{eq:3.4} reads
\cite{Dias:2007}
\begin{subequations}
\label{eq:4.1}
\begin{align}
Q(x,p,t;\hbar) & = \sec^2(\hbar t) x \exp \left( \frac{2}{\hbar} \tan(\hbar t)
    xp \right),
\label{eq:4.1a} \displaybreak[0] \\
P(x,p,t;\hbar) & = \sec^2(\hbar t) p \exp \left( -\frac{2}{\hbar} \tan(\hbar t)
    xp \right),
\label{eq:4.1b}
\end{align}
\end{subequations}
for $\abs{t} < \frac{\pi}{2\hbar}$. This solution is a deformation of a
classical one given by the limit $\hbar \to 0$
\begin{equation*}
Q_C(x,p,t) = x e^{2t xp}, \quad P_C(x,p,t) = p e^{-2t xp}.
\end{equation*}
The induced quantum flow $\Phi_t$ is an example of a flow for
which $\Phi_t$, for every $t \in
(-\frac{\pi}{2\hbar},\frac{\pi}{2\hbar}) \setminus \{0\}$, is not
a classical symplectomorphism, since
\begin{equation*}
\{Q(t),P(t)\} = \sec^4(\hbar t) \neq 1.
\end{equation*}

In accordance with \eqref{eq:3.1} the quantum flow $\Phi_t$ transforms the Moyal
product to the following product
\begin{equation*}
f \star_t g = f \exp \left( \frac{1}{2} i\hbar \overleftarrow{D_{x}}
    \overrightarrow{D_{p}} - \frac{1}{2} i\hbar \overleftarrow{D_{p}}
    \overrightarrow{D_{x}} \right) g,
\end{equation*}
where
\begin{align*}
D_x & = \sec^2(\hbar t) \bigl(1 + 2t a(\hbar t) xp\bigr)
    \exp\bigl(2t a(\hbar t) xp\bigr) \partial_x
    - 2t \sec^2(\hbar t) a(\hbar t) p^2 \exp\bigl(2t a(\hbar t) xp\bigr)
    \partial_p, \\
D_p & = 2t \sec^2(\hbar t) a(\hbar t) x^2 \exp\bigl(-2t a(\hbar t) xp\bigr)
    \partial_x
    + \sec^2(\hbar t) \bigl(1 - 2t a(\hbar t) xp\bigr)
    \exp\bigl(-2t a(\hbar t) xp\bigr) \partial_p,
\end{align*}
and $a(x) = \frac{\tan(x)}{x\sec^4(x)}$. Moreover, the isomorphism $S_t$
associated with $\Phi_t$ and intertwining the Moyal product with the
\nbr{\star_t}product, up to the second order in $\hbar$, takes the form
\begin{align}
S_t & = 1 + \hbar^2 \biggl( \frac{1}{6}(3t^2 x^3 + 4t^3 x^4 p) \partial_x^3
+ \frac{1}{6}(3t^2 p^3 - 4t^3 xp^4) \partial_p^3
+ \frac{1}{2}(-tp - t^2 xp^2 + 4t^3 x^2p^3) \partial_x \partial_p^2 \nonumber \\
& \quad {} + \frac{1}{2}(tx - t^2 x^2 p - 4t^3 x^3 p^2)
\partial_x^2 \partial_p + (2t^2 x^2 + 2t^3 x^3 p) \partial_x^2 +
(2t^2 p^2 - 2t^3 x p^3) \partial_p^2 + (-2t^2 xp)
\partial_x \partial_p \biggr) + o(\hbar^4).
\label{eq:4.2}
\end{align}
Indeed, expanding relations \eqref{eq:3.2} with respect to $\hbar$ one can
prove that $S_t$ in the above form satisfies these relations up to $o(\hbar^2)$.

From the fact that $\Phi_t$ is a purely quantum trajectory, we deal with the
quantum group multiplication \eqref{eq:3.15} for $\{\Phi_t\}$ as well as the quantum action
\eqref{eq:3.7} of $\Phi_t$ on observables and states. Indeed, expanding \eqref{eq:4.1} with respect to $\hbar$:
\begin{align*}
Q(x,p,t;\hbar) & = Q_C \left(1 + \hbar^2 \left(t^2 + \frac{2}{3}t^3 xp\right)
    \right) + o(\hbar^4), \displaybreak[0] \\
P(x,p,t;\hbar) & = P_C \left(1 + \hbar^2 \left(t^2 - \frac{2}{3}t^3 xp\right)
    \right) + o(\hbar^4)
\end{align*}
and applying isomorphism $S_t$ \eqref{eq:4.2}, the quantum composition law
\begin{align*}
Q(t_1 + t_2) & = S_{t_2}Q(t_1) \circ \Phi_{t_2} = S_{t_1}Q(t_2) \circ
    \Phi_{t_1}, \\
P(t_1 + t_2) & = S_{t_2}P(t_1) \circ \Phi_{t_2}= S_{t_1}P(t_2) \circ
    \Phi_{t_1}
\end{align*}
holds up to $o(\hbar^2)$. Note also, that the flow $\Phi_t$ is not
defined for all $t \in \mathbb{R}$ but only on an interval
$(-\frac{\pi}{2\hbar},\frac{\pi}{2\hbar})$, contrary to classical
flows which are always globally defined. This is an interesting
result showing that in general the quantum time evolution do not
have to be defined for all instances of time $t$.



\begin{thebibliography}{23}
\expandafter\ifx\csname natexlab\endcsname\relax\def\natexlab#1{#1}\fi
\providecommand{\bibinfo}[2]{#2}
\ifx\xfnm\relax \def\xfnm[#1]{\unskip,\space#1}\fi
\bibitem[{Dirac(1958)}]{Dirac:1958}
\bibinfo{author}{P.~A.~M. Dirac}, \bibinfo{title}{The Principles of Quantum
  Mechanics}, \bibinfo{publisher}{Oxford University Press},
  \bibinfo{address}{New York}, \bibinfo{edition}{fourth} edition,
  \bibinfo{year}{1958}.
\bibitem[{de~Broglie(1926)}]{Broglie:1926a}
\bibinfo{author}{L.~de~Broglie}, \bibinfo{journal}{C. R. Acad. Sci.}
  \bibinfo{volume}{183} (\bibinfo{year}{1926}) \bibinfo{pages}{272}.
\bibitem[{de~Broglie(1927)}]{Broglie:1927b}
\bibinfo{author}{L.~de~Broglie}, \bibinfo{journal}{J. Phys. Radium}
  \bibinfo{volume}{8} (\bibinfo{year}{1927}) \bibinfo{pages}{225}.
\bibitem[{Bohm(1952{\natexlab{a}})}]{Bohm:1952a}
\bibinfo{author}{D.~Bohm}, \bibinfo{journal}{Phys. Rev.} \bibinfo{volume}{85}
  (\bibinfo{year}{1952}{\natexlab{a}}) \bibinfo{pages}{166--179}.
\bibitem[{Bohm(1952{\natexlab{b}})}]{Bohm:1952b}
\bibinfo{author}{D.~Bohm}, \bibinfo{journal}{Phys. Rev.} \bibinfo{volume}{85}
  (\bibinfo{year}{1952}{\natexlab{b}}) \bibinfo{pages}{180--193}.
\bibitem[{Holland(1993)}]{Holland:1993}
\bibinfo{author}{P.~R. Holland}, \bibinfo{title}{The Quantum Theory of Motion:
  An Account of the de {B}roglie-{B}ohm Causal Interpretation of Quantum
  Mechanics}, \bibinfo{publisher}{Cambridge University Press},
  \bibinfo{address}{Cambridge}, \bibinfo{year}{1993}.
\bibitem[{Wyatt(2005)}]{Wyatt:2005}
\bibinfo{author}{R.~E. Wyatt}, \bibinfo{title}{Quantum Dynamics with
  Trajectories: Introduction to Quantum Hydrodynamics},
  \bibinfo{publisher}{Springer}, \bibinfo{address}{Berlin},
  \bibinfo{year}{2005}.
\bibitem[{Littlejohn(1986)}]{Littlejohn:1986}
\bibinfo{author}{R.~G. Littlejohn}, \bibinfo{journal}{Phys. Rep.}
  \bibinfo{volume}{138} (\bibinfo{year}{1986}) \bibinfo{pages}{193--291}.
\bibitem[{Dias and Prata(2007)}]{Dias:2007}
\bibinfo{author}{N.~C. Dias}, \bibinfo{author}{J.~N. Prata},
  \bibinfo{journal}{J. Math. Phys.} \bibinfo{volume}{48} (\bibinfo{year}{2007})
  \bibinfo{pages}{012109}.
\bibitem[{Krivoruchenko and Faessler(2007)}]{Krivoruchenko:2007}
\bibinfo{author}{M.~I. Krivoruchenko}, \bibinfo{author}{A.~Faessler},
  \bibinfo{journal}{J. Math. Phys.} \bibinfo{volume}{48} (\bibinfo{year}{2007})
  \bibinfo{pages}{052107}.
\bibitem[{Rieffel(1997)}]{Rieffel:1997}
\bibinfo{author}{M.~A. Rieffel}, \bibinfo{journal}{Can. J. Math.}
  \bibinfo{volume}{49} (\bibinfo{year}{1997}) \bibinfo{pages}{160--174}.
\bibitem[{Osborn and Molzahn(1995)}]{Osborn:1995}
\bibinfo{author}{T.~A. Osborn}, \bibinfo{author}{F.~H. Molzahn},
  \bibinfo{journal}{Ann. Phys.} \bibinfo{volume}{241} (\bibinfo{year}{1995})
  \bibinfo{pages}{79--127}.
\bibitem[{Bayen et~al.(1978{\natexlab{a}})Bayen, Flato, Fr{\o}nsdal,
  Lichnerowicz, and Sternheimer}]{Bayen:1978a}
\bibinfo{author}{F.~Bayen}, \bibinfo{author}{M.~Flato},
  \bibinfo{author}{C.~Fr{\o}nsdal}, \bibinfo{author}{A.~Lichnerowicz},
  \bibinfo{author}{D.~Sternheimer}, \bibinfo{journal}{Ann. Phys.}
  \bibinfo{volume}{111} (\bibinfo{year}{1978}{\natexlab{a}})
  \bibinfo{pages}{61--110}.
\bibitem[{Bayen et~al.(1978{\natexlab{b}})Bayen, Flato, Fr{\o}nsdal,
  Lichnerowicz, and Sternheimer}]{Bayen:1978b}
\bibinfo{author}{F.~Bayen}, \bibinfo{author}{M.~Flato},
  \bibinfo{author}{C.~Fr{\o}nsdal}, \bibinfo{author}{A.~Lichnerowicz},
  \bibinfo{author}{D.~Sternheimer}, \bibinfo{journal}{Ann. Phys.}
  \bibinfo{volume}{111} (\bibinfo{year}{1978}{\natexlab{b}})
  \bibinfo{pages}{111--151}.
\bibitem[{Dito and Sternheimer(2002)}]{Dito.Sternheimer:2002}
\bibinfo{author}{G.~Dito}, \bibinfo{author}{D.~Sternheimer}, in:
  \bibinfo{editor}{G.~Halbout} (Ed.), \bibinfo{booktitle}{Deformation
  quantization}, volume~\bibinfo{volume}{1} of \textit{\bibinfo{series}{{IRMA}
  lectures in mathematics and theoretical physics}}, \bibinfo{publisher}{Walter
  de Gruyter}, \bibinfo{address}{Berlin, New York}, \bibinfo{year}{2002}, pp.
  \bibinfo{pages}{9--54}.
\bibitem[{Gutt(2000)}]{Gutt:2000}
\bibinfo{author}{S.~Gutt}, in: \bibinfo{editor}{G.~Dito},
  \bibinfo{editor}{D.~Sternheimer} (Eds.), \bibinfo{booktitle}{Conf{\'e}rence
  {M}osh{\'e} {F}lato 1999: quantization, deformations, and symmetries},
  volume~\bibinfo{volume}{21} of \textit{\bibinfo{series}{Mathematical physics
  studies}}, \bibinfo{publisher}{Kluwer Academic Publishers},
  \bibinfo{address}{Netherlands}, \bibinfo{year}{2000}, pp.
  \bibinfo{pages}{217--254}.
\bibitem[{Weinstein(1994)}]{Weinstein:1994b}
\bibinfo{author}{A.~Weinstein}, in: \bibinfo{booktitle}{S{\'e}minaire
  Bourbaki}, volume~\bibinfo{volume}{36}, \bibinfo{publisher}{Association des
  Collaborateurs de Nicolas Bourbaki}, \bibinfo{year}{1993--1994}, pp.
  \bibinfo{pages}{389--409}.
\bibitem[{B{\l}aszak and Doma{\'n}ski(2012)}]{Blaszak:2012}
\bibinfo{author}{M.~B{\l}aszak}, \bibinfo{author}{Z.~Doma{\'n}ski},
  \bibinfo{journal}{Ann. Phys.} \bibinfo{volume}{327} (\bibinfo{year}{2012})
  \bibinfo{pages}{167--211}.
\bibitem[{Bordemann and Waldmann(1998)}]{Bordemann:1998}
\bibinfo{author}{M.~Bordemann}, \bibinfo{author}{S.~Waldmann},
  \bibinfo{journal}{Commun. Math. Phys.} \bibinfo{volume}{195}
  (\bibinfo{year}{1998}) \bibinfo{pages}{549--583}.
\bibitem[{Waldmann(2005)}]{Waldmann:2005}
\bibinfo{author}{S.~Waldmann}, \bibinfo{journal}{Rev. Math. Phys.}
  \bibinfo{volume}{17} (\bibinfo{year}{2005}) \bibinfo{pages}{15--75}.
\bibitem[{Dias and Prata(2004)}]{Dias:2004b}
\bibinfo{author}{N.~C. Dias}, \bibinfo{author}{J.~N. Prata},
  \bibinfo{journal}{Ann. Phys.} \bibinfo{volume}{313} (\bibinfo{year}{2004})
  \bibinfo{pages}{110--146}.
\bibitem[{Dias and Prata(2001)}]{Dias:2001}
\bibinfo{author}{N.~C. Dias}, \bibinfo{author}{J.~N. Prata},
  \bibinfo{journal}{J. Math. Phys.} \bibinfo{volume}{42} (\bibinfo{year}{2001})
  \bibinfo{pages}{5565--5579}.
\bibitem[{Dias and Prata(2004)}]{Dias:2004}
\bibinfo{author}{N.~C. Dias}, \bibinfo{author}{J.~N. Prata},
  \bibinfo{journal}{J. Math. Phys.} \bibinfo{volume}{45} (\bibinfo{year}{2004})
  \bibinfo{pages}{887--901}.
\bibitem[{B{\l}aszak and Doma{\'n}ski(2012)}]{Blaszak:2012c}
\bibinfo{author}{M. B{\l}aszak}, \bibinfo{author}{Z. Doma{\'n}ski},
  \bibinfo{note}{eprint arXiv:1208.2835 [math-ph]} (\bibinfo{year}{2012}).

\end{thebibliography}
\end{document}